\documentstyle[12pt]{article}
\def\pictures{y }
\ifx\pansw\pictures
\input epsf
\input prepictex
\input pictex
\input postpictex
\else
\fi
\newdimen\tdim
\newcommand{\mysection}[1]{\setcounter{equation}{0}\section{#1}}

\begin{document}


\newcommand{\nc}{\newcommand}
\nc{\beq}{\begin{equation}}     \nc{\eeq}{\end{equation}}
\nc{\beqa}{\begin{eqnarray}}    \nc{\eeqa}{\end{eqnarray}}
\nc{\lsim}{\begin{array}{c}\,\sim\vspace{-21pt}\\< \end{array}}
\nc{\gsim}{\begin{array}{c}\sim\vspace{-21pt}\\> \end{array}}
\nc{\create}{\hat{a}^\dagger}   \nc{\destroy}{\hat{a}}
\nc{\kvec}{\vec{k}}             \nc{\kvecp}{\vec{k}^\prime}
\nc{\kvecpp}{\vec{k}^{\prime\prime} }   \nc{\kb}{\bf k}
\nc{\kbp}{{\bf k}^\prime}       \nc{\kbpp}{{\bf k}^{\prime\prime} }
\nc{\bfk}{{\bf k}}              \nc{\cohak}{a_{{\bf k}}}
\nc{\cohap}{a_{{\bf p}}}        \nc{\cohaq}{a_{{\bf q}}}
\nc{\cohbk}{b^*_{{\bf k}}}      \nc{\cohbp}{b^*_{{\bf p}}}
\nc{\cohbq}{b^*_{{\bf q}}}      \nc{\bark}{\bar{k}}

\begin{titlepage}
\begin{center}
\setlength{\baselineskip}{0.20in}
{\hbox to\hsize{March 1995\hfill
hep-ph/9502276, DUKE-TH-95-85}}
{\hbox to\hsize{\hfill  JHU-TIPAC-950004, YCTP-P4-95, EFI-95-06}}
\setlength{\baselineskip}{0.35in}
\bigskip
{\Large \bf Wavepacket Dynamics in Yang--Mills Theory } \\
\vskip 0.25in

{\bf C.R. Hu}\footnotemark[1],
{\bf S.G. Matinyan}\footnotemark[2],
{\bf B. M\"uller}\footnotemark[3], and
{\bf A. Trayanov}\footnotemark[4]\\
{\it Department of Physics, Duke University, Durham NC 27708-0305} \\[.05in]
{\bf T.M. Gould\footnotemark[5]}\\
{\it Dept. of Physics and Astronomy, The Johns Hopkins University,
Baltimore MD 21218}\\[.05in]
{\bf S.D.H. Hsu}\footnotemark[6]\\
{\it Sloane Physics Laboratory, Yale University,
New Haven CT 06511} \\[.05in]
{\bf E.R. Poppitz}\footnotemark[7]\\
{\it The Enrico Fermi Institute, University of Chicago,
Chicago IL 60637}\\[.15in]

\smallskip
{\bf Abstract}\\[-0.25in]

\begin{quote}
\setlength{\baselineskip}{0.2in}
We discuss the results of numerical simulations of colliding
wavepackets in $SU(2)$ Yang--Mills theory. We investigate their
behavior as a function of amplitude and momentum distribution.
We find regions in our parameter space in which initial wave
packets scatter into final configurations with dramatically
different momentum distributions.  These results constitute
new classical trajectories with multiparticle boundary conditions.
We explain their relevance for the calculation of scattering
amplitudes in the semiclassical approximation.  Finally, we give
directions for future work.
\end{quote}
\end{center}

\footnotetext[1]{chaoran@phy.duke.edu}
\footnotetext[2]{On leave from Yerevan Physical Institute, Yerevan 375036,
Armenia. E-mail: ani@phy.duke.edu}
\footnotetext[3]{muller@phy.duke.edu}
\footnotetext[4]{Permanent address: NCSC, Research Triangle Park, North
Carolina 27708. E-mail: nasco@phy.duke.edu}
\footnotetext[5]{gould@fermi.pha.jhu.edu}
\footnotetext[6]{hsu@hsunext.physics.yale.edu}
\footnotetext[7]{epoppitz@yukawa.uchicago.edu}

\end{titlepage}

\renewcommand{\thepage}{\arabic{page}}
\setcounter{page}{1}
\mysection{Introduction }

Two recent observations on the dynamics of Yang--Mills theory can be
combined to yield new nonperturbative information about scattering
amplitudes in non-Abelian gauge theories.  In \cite{GHP}, it was shown
that the existence of real trajectories satisfying multiparticle
boundary conditions implies high multiplicity scattering amplitudes
which are not suppressed in the weak coupling limit.  In \cite{GMMT},
it was found that a standing plane wave is unstable in $SU(2)$ Yang-Mills
theory.

The authors of \cite{GHP} developed a stationary phase approximation for
computing scattering amplitudes.  When the stationary  trajectory is real,
it satisfies an initial value problem, consisting of the classical field
equations with wavepacket initial conditions, and describes a classically
allowed transition~\footnote{The general case of {\it complex} trajectories
is discussed in \cite{GHP}.}. In this case, there is no exponential
suppression of the associated scattering amplitude at small coupling, as
opposed to disallowed, or tunneling transitions.  The first step in this
program is to find such trajectories.  The complicated nature of the field
equations necessitates a computational approach.

Indeed, progress in this direction has already been made.  In studying the
chaotic dynamics of Yang--Mills theory, the authors of \cite{GMMT} have
found evidence for an instability in the field equations which leads to a
growth of long wavelength modes from initial short wavelength standing
plane waves.  The mechanism finds its origin in the non-Abelian spin-field
coupling, which drives an instability for  perturbations with isospin
polarizations orthogonal to the standing plane wave~\cite{GMMT}.

This implies the existence of classical trajectories of the type required
by \cite{GHP}, if the instability persists in the regime of localized
wavepackets.  One purpose of the present investigation is to determine the
extent to which this is the case.  The authors of \cite{GHP} have already
identified a mechanism which may produce interesting classical trajectories.
Our goal in the current paper is to investigate this mechanism through
numerical simulations on a lattice.  For simplicity, we have studied
wavepackets with constant profiles in the transverse directions.

Our interest in the problem addressed in this paper has its origin in the
work of \cite{RE}. These authors suggested that multiparticle production
could be unsuppressed in electroweak interactions at high energies
and could manifest itself in fermion number violating processes.  In order
to investigate this phenomena further, previous authors have studied the
classical analog of high energy particle collisions~\cite{RJTR,GNV}.  The
studies of one dimensional Abelian Higgs gauge theory~\cite{RJTR} and
$\lambda\phi^4$ theory~\cite{GNV} have shown no indication for the
classical analog of multiparticle scattering.  The work of \cite{GHP}
establishes a precise connection between the  classical results of
\cite{RJTR,GNV} and quantum scattering amplitudes.

This paper is organized as follows.  In Section 2, we describe the
numerical simulation.  In Section 3, we discuss some of our results and
their implications.  In Section 4, we explain the relevance of our
computations to realistic three dimensional scattering.  In Section 5,
we summarize and give directions for future research.

\mysection{Classical Yang--Mills Theory on the Lattice}

In this section, we describe the numerical simulation~\cite{GT}.  For
simplicity, we work with a one dimensional lattice with physical size
$L = N a$, where $N$ is the number of lattice sites and $a$ the lattice
spacing. We arrange initially two Gaussian wavepackets with average momenta
$\left(\bark,0,0,\pm\bark\right)$,
and width $\Delta k$, between the maximum and minimum longitudinal momenta
on the lattice $k_{min}, k_{max}$.
\begin{equation}
\left(\, k_{min} = 2 \pi/Na\,\right)~\ll~
\Delta k~\ll~\bark~\ll~
\left(\, k_{max} = 2 \pi/2a\,\right) ~.
\end{equation}
The packets are fairly narrow in momentum space, $\Delta k/\bark \ll 1$.

The wavepackets are constant in the transverse $x, y$ directions.
Since, at time $t = 0$, the two wavepackets are arranged to be well--separated
from each other and from the lattice boundaries, the gauge field $A^{c,i}$ is
initially
\begin{equation}
\label{initial}
A^{c,i} ~=~ A_R^{c,i} ~+~ A_L^{c,i}  ~.
\end{equation}
We work in the $A^0 = 0$ gauge, $c$ is the isospin index, $i$ the Lorentz
index.  $A_R$ is a right--moving wavepacket, initially centered at $z=Z_R$.
$A_L$ is a left--moving wavepacket, initially centered at $z=Z_L$.
In practice, $Z_R$ and $Z_L$ are chosen so that they are symmetrical about
the center of the longitudinal lattice and $\Delta Z \equiv Z_L - Z_R$
is large compared to $1/\Delta k$.

Both wavepackets are chosen to be polarized in the same spatial direction
($i = 2$) for simplicity.  Furthermore, the right--moving wavepacket is
polarized in an isospin direction defined by a unit vector $\hat{n}_R =
(0,0,1)$
\begin{equation}
A_R^{c,i} ~=~ \delta^{i2}~\hat{n}_R^c~\phi\left(\, z-Z_R, -t\,\right)~,
\end{equation}
and the left--moving wavepacket is polarized in an isospin direction defined
by another unit vector $\hat{n}_L$
\begin{equation}
A_L^{c,i} ~=~ \delta^{i2}~\hat{n}_L^c~\phi\left(\, z - Z_L, t\,\right)~.
\end{equation}
Each wavepacket would therefore be stable in the absence of the other
and would propagate as a collection of free travelling waves.  The relative
rotation between the two isospin vectors, $\hat{n}_R\cdot\hat{n}_L \equiv
\cos\theta_c$, is a free parameter in our computation.  Nonlinear terms in
the Yang--Mills equations arise from commutators which vanish at
$\theta_c=0$.

The wavepacket is chosen as a Gaussian
\begin{equation}
\label{wp}
\phi\left(z,t\right) ~=~
\sqrt{2\,\Delta k \over \pi^{1/2}\,\sigma\,\bark}~
\exp{\left(\, - \frac{1}{2}(\Delta k)^2 \left(z + t\right)^2 \,\right)}\,
\cos \left(\, \bark \left( z + t\right) \,\right) ~.
\end{equation}
This expression comes from the Fourier transform of a Gaussian momentum
space distribution, after dropping some exponentially small corrections
${\cal O}(e^{-(\bark / \Delta k)^2 / 2})$.  There is no intrinsic mass
scale in classical Yang--Mills theory.  So, dimensionful parameters will
be determined by the lattice spacing $a$.

Our study is based on the Hamiltonian formulation of lattice $SU(2)$
gauge theory~\cite{CRUK}, in which the dynamic variables are link variables
defined as
\begin{equation}
U_l ~=~ \exp{\left(\, -\frac{i}{2}\, g\, a\, A_l^a\,\tau^a\,\right)}~.
\end{equation}
where $l$ stands for the link index and $\tau^c$ ($c=1, 2, 3$) are
Pauli matrices.

For $z < L/2$, the link variable is approximately that of the right--moving
wavepacket:
\begin{equation}
\label{linkright}
U_l ~=~
\cos \left(\,
\frac{1}{2}\, g\, a\,\phi\left(\, z-Z_R, -t\,\right)\,\right) ~+~
i\,\tau\cdot\hat{n}_R\,
\sin \left(\,
\frac{1}{2}\, g\, a\,\phi\left(\, z-Z_R, -t\,\right)\,\right)
\hspace{0.5cm} {\rm if} \hspace{5mm} l = 2 ~.
\end{equation}
For $z > L/2$, the link variable is approximately that
of the left--moving wavepacket:
\begin{equation}
\label{linkleft}
U_l ~=~
\cos \left(\,
\frac{1}{2}\, g\, a\,\phi\left(\, z-Z_L, t\,\right)\,\right) ~+~
i\,\tau\cdot\hat{n}_L\,
\sin \left(\,
\frac{1}{2}\, g\, a\,\phi\left(\, z-Z_L, t\,\right)\,\right)
\hspace{0.5cm} {\rm if} \hspace{5mm} l = 2 ~.
\end{equation}
We are evolving wavepackets which are initially free, and therefore
the initial
value of $\dot{U}_l$ is just the time derivative
of $U_l$ as given in (\ref{linkright}) and (\ref{linkleft}).

Our ansatz for the initial wavepackets has four free parameters.  The
relative isospin polarizations of the two packets is parametrized by
$\theta_c$.  There are two free momentum scales: the average momentum of
the wavepackets $\bark$ and their width $\Delta k$.  Finally, the amplitude
is controlled by the free parameter $\sigma$.  The coupling $g$ can be
rewritten in terms of the parameter $\sigma$ by rescaling the field in the
equations and initial conditions. To see this explicitly, let $A' =
\sqrt{\sigma} A$
and $F'^{\mu\nu}=\sqrt{\sigma} F^{\mu\nu}=\partial^{\mu}A'^{\nu}-
\partial^{\nu} A'^{\mu}+ig'[A'^{\mu},~A'^{\nu}]$, where $g' = g/\sqrt{\sigma}$.
Then, with respect to the scaled gauge potential $A'^{\mu}$,
the equation of motion reads
\begin{equation}
\label{scaling}
\partial_{\mu}F'^{\mu\nu} ~+~ ig'\,\left[A'_{\mu}, ~ F'^{\mu\nu}\right] ~=~
0 ~.
\end{equation}
Since $A'^{\mu}$ is only dynamically dependent on the coupling $g'$,
we conclude that the system shows the same dynamics for different values of
$g$ and $\sigma$, as long as the ratio $g'=g/\sqrt{\sigma}$ is held fixed.

In the next section, we describe the behavior of our simulations with
respect to the parameters mentioned above (we fix the coupling $g = 1$).

\newpage
\mysection{Results from Colliding Wavepackets}

{\bf 3.1}~~{\em Snapshots}
\smallskip

First, we present two figures of the field amplitude itself to illustrate
some collisions qualitatively, plotting successive time slices of the norm
of the field.  Figure 1 shows an elastic collision of wavepackets with no
relative isospin polarization.  The wavepackets pass through each other
without interacting, as expected.  This constitutes a check on our
numerical procedure and shows the nonlinearities introduced by the
formulation in terms of compact lattice gauge fields do not affect the
scattering process.

Figure 2 illustrates an inelastic collision of wavepackets with maximal
(orthogonal) relative isospin polarization, for the same values of
$\bark$ and $\Delta k$.  The wavepackets are distorted after the collision
and clearly contain modes in a range around the initial momentum distribution.
As we will discuss below, inelastic collisions remain qualitatively similar
for relative polarizations as small as ${\cal O}\left(10^{-12}\right)$, all
other parameters held fixed.  In the following two subsections, we
specialize to the case of {\em orthogonal} isospin polarization.

It should be noted that the fields do not linearize at late times in our
simulation.  In the absence of dispersion, the energy density of each packet
is always bounded in time in one dimensional collisions, and the fields are
not expected to linearize at large times.  As a result, quantities like
energy spectra do not settle down to constants at late times.  One
consequence for us is that it will not make sense to infer any
characteristics about  a ``final'' state from the late time behavior on
our lattice.  In the full three dimensional situation, geometry allows energy
densities to spread out in space and to approach the linear regime.  In this
case, we expect to see the spectrum approach a constant at late times.
Lacking this time scale in our one dimensional simulation, we simply allow
our simulation to run as long as it would take two non--interacting
wavepackets to pass through each other and become well--separated,
$t\simeq 600$, see Figure 1.  The relevance of our one dimensional
simulations for the three dimensional physics will be discussed further
in section 4.

\smallskip
{\bf 3.2}~~{\em Energy spectrum}
\smallskip

To illustrate the broadening of momentum distribution after collision,
Figures 3 and 4, in correspondence with Figures 1 and 2 respectively,
show the time evolution of the absolute value of the Fourier transform of
the energy density
\begin{equation}
\label{energy}
{\cal E}\left({\bf k}, t\right) ~=~
\frac{1}{4}\,\left|\,\int {d^3x}~e^{i{\bf k}\cdot{\bf x}}~{\rm Tr}~
\left[\, {\bf E}^2\left({\bf x}, t\right) ~+~
{\bf B}^2\left({\bf x}, t\right) \,\right]\,\right|~,
\end{equation}
where ${\bf E}$ is the isospin electric field and ${\bf B}$ the isospin
magnetic
field.  As expected, the spectrum for parallel polarization (Figure 3) remains
unchanged in the time allowed by the simulation, with a peak at roughly
twice the average momentum $\bar k$.  The spectrum for orthogonal
polarization (Figure 4) is strongly disrupted due to nonlinear interactions.
In each of these figures, the spike at $k=0$ is the total energy of the
wavepackets, and its constancy provides a check on our computation.

\smallskip
{\bf 3.3}~~{\em Amplitude and width dependence}
\smallskip

By rescaling coordinates,
the  initial configuration (\ref{wp}) can be rewritten
\begin{equation}
\label{scaled}
\phi'(z', t') ~\equiv~ \phi(z, t)/\bar k ~=~ \sqrt{2R\over \pi^{1/2}}~
\exp{\left(\, - \left(z' + t'\right)^2/2\, W^2 \,\right)}\,
\cos \left( z' + t'\right) ~,
\end{equation}
in terms of dimensionless spacetime coordinates,
$z'\equiv\bar k z$ and  $t'\equiv\bar k t$, and dimensionless parameters,
$R \equiv \Delta k/\sigma\bar k^3$ and $W\equiv \bar k/\Delta k$.
We refer to $R$ as the amplitude parameter and $W$ as the
width parameter.
The initial separation between the wavepackets in dimensionless units
is $\bar k \Delta Z$.
Thus,  the time evolution
of configurations with the same dimensionless parameters are
identical in the {\em continuum}.

However,
the lattice simulation necessarily involves two more dimensional parameters:
the length $L$ of the lattice and the lattice spacing $a$.
We have chosen both $L$ and also $\Delta Z$ large enough compared to the
width of the wavepacket that their effects can be neglected.
The scaled lattice spacing $a'=\bar k a$  is the number of lattice points
representing a single wavelength, and specifies how close the
simulation is to the continuum limit.
We have checked that the qualitative behavior of our simulation is not
sensitive to the choice of $a'$.
Therefore,
we  present our results with fixed lattice spacing $a=1$ and $\bar k=\pi/10$.
(This corresponds to $20$ lattice points in a single wavelength).
The scaling property explicit in (\ref{scaled}) provides a check that
our lattice is close to the continuum limit.
Simulations with the same dimensionless parameters, $R$ and $W$,
should be identical.  We have verified that our results are consistent with
this scaling.

Figure 5 displays the dependence of the energy spectrum on the initial
dimensionless amplitude $R$, with the width parameter $W$ fixed.
The nonlinearity clearly gets weaker as the amplitude $R$
is decreased.
Larger values of $\sigma$ require longer times for
the development of comparable nonlinearities.
This is analogous to the case of the standing wave configuration
where the largest Lyapunov exponent is a linear function of
the amplitude~\cite{GMMT}.

Figures 6 and 7 display the dependence of the energy spectrum on
the width parameter $W$, with $R$ fixed.
In the figures, $W$ varies by a factor of $5$.
We expect that wider packets overlap for longer times during collision
and should behave more nonlinearly.
We see that the instability is always more pronounced in the widest
packet ($W = 25$) versus the narrowest ($W = 5$).
However, lacking a quantitative measure of nonlinearity it is hard
to characterize exactly what is happening when we compare other
values of $W$.

\smallskip
{\bf 3.4}~~{\em Dependence on isospin polarization}
\smallskip

We now consider the dependence on the initial isospin polarization of the
wavepackets.  As given in Section 2, the right--moving packet is initially
polarized in the isospin $c=3$ direction, while the left--moving packet is
polarized in the isospin $c=2,3$ plane.  The relative angle $\theta_c$ is
varied from $0$ to $\pi$.

Figure 8 shows the spectra at $t=600$ for different polarizations.
Figure 9 shows the evolution of energy contained in the isospin $c=1$
component, modes orthogonal to both initial packets.  An exponential
increase is seen for time $t \approx 200$ or later in Figure 9.  The
largest overlap between the two wavepackets occurs at time $t \approx 200$
when the centers of the two wavepackets meet.  However, the energy drops
after an exponential growth.  Again, it does not approach a constant at
late times after the collision due to the absence of dispersion in the
one dimensional simulation.

It is remarkable that the nonlinear behavior persists until very small
relative polarizations, near $\theta_c = 10^{-12}\pi$, where one might
expect nonlinear effects to be perturbatively small.  This was the case
in the limit of large $\sigma$, where nonlinear effects disappeared
entirely for $\sigma$ as small as ${\cal O}(10)$ (see Figure 5).
However, the dynamics respond quite differently in the two limits.  For
large $\sigma$, the initial amplitude is small, ${\cal O}(\sigma^{-1/2})$,
and all nonlinear terms in the field equations are suppressed.
For small $\theta_c$ however, the initial amplitude is not suppressed,
${\cal O}(\theta_c^0)$, and can drive instabilities.

Recall the initial gauge potential (\ref{initial}) has the form
\begin{eqnarray}
\label{initx}
A^i(t, \vec{x}) &=& \delta^{i2}~\left(\,
\phi_R ~+~ \phi_L~\cos\theta_c\,\right)~\tau^3 ~+~
\delta^{i2}~\phi_L~\sin\theta_c~\tau^2 ~, \\
&=&
\delta^{i2}~
\left(\,\phi_R ~+~ \phi_L\,\right)~\tau^3 ~+~
\delta^{i2}~\phi_L~\theta_c~\tau^2 ~+~ {\cal O}(\theta_c^2) ~.   \nonumber
\end{eqnarray}
It evolves at some short time $t$ later to the more general form
\begin{eqnarray}
\label{pert}
A^i(t, \vec{x}) ~=~ \delta^{i2}
\left(\,\phi_R ~+~ \phi_L\,\right)~\tau^3 ~+~
\delta^{i2}~\phi_L~\theta_c~\tau^2 ~+~ b^a_i~\tau^a ~+~
{\cal O}(\theta_c^2) ~,
\end{eqnarray}
where $b^a_i$ are small perturbations and may have any polarizations.
The background field (\ref{initx}) driving this perturbation
is not dissimilar from that studied in \cite{GMMT}.
In the limit $\theta_c\approx 0$,
it is a momentum superposition of isospin--polarized standing waves,
which is unstable to small isospin--orthogonal perturbations,
just as a single polarized standing wave is~\cite{GMMT}.
The Yang--Mills equations for (\ref{pert}) then imply equations for
the perturbation of the form
\begin{equation}
\label{b}
\partial^\mu \partial_{\left[\mu\right.} b_{\left.\nu\right]} ~+~
{\cal O}\left(b\cdot\left(\,\phi_R ~+~ \phi_L\,\right)\right) ~+~
{\cal O}(b^2) ~+~ {\cal O}(\theta_c)~=~ 0 ~.
\end{equation}
in which the coupling of the perturbation to the background
is ${\cal O}(\theta_c^0)$.
This coupling drives the instability (at least initially) with
a growth rate determined by the associated largest Lyapunov
exponent~\cite{GMMT}.
Thus, one understands why the instability persists even for
very small relative polarizations, as seen in Figure 8.
As a further check on our understanding of this limit,
the cases with very small $\theta_c$ in Figures 8 and 9
display the similar exponential growth found in \cite{GMMT}.

\mysection{Implications for three dimensions}

In this section we discuss the implications of our one dimensional
simulation for the behavior of wavepackets in three dimensions. Obviously,
our simulation applies directly to three dimensional configurations which
are translationally invariant in the $x-y$ plane. However, any configuration
of this type must contain an infinite number of particles if the gauge field
has nonzero amplitude.

Our simulation also applies to initial wavepackets with finite but large
$x-y$ extent.  As sketched in Figure 10, the evolution of the core region
$x,y \approx 0$ is represented by the dotted part until time $t \approx
r_{\perp}$ in the world plane, while the evolution of the edge effect is
bounded by two solid lines starting at $r = r_{\perp}$ and $t = 0$. Clearly,
if the transverse extent of the wavepackets $r_{\perp}$ is larger than
the total elapsed time of the simulation,
\begin{equation}
\label{causality}
r_{\perp} > T_s ~,
\end{equation}
then causality guarantees that the three dimensional evolution in the core
region will be the same as that found in our simulation. Of course, we cannot
deduce from our simulation what the final three dimensional configuration
will be.  However, we can at least conclude that some nonlinear activity,
such as the redistribution of energy in momentum space, has occurred.

Thus our numerical results are predictive of the behavior of three
dimensional wavepackets with pancake--like geometry (see Figure 11).
This is precisely the geometry that one would expect for an initially
roughly spherical wavepacket which has been boosted relative to the lab
frame.  For such wavepackets we can relate the parameters describing our
initial configurations to the particle number of corresponding three
dimensional wavepackets. Our choice of normalization in (\ref{wp}) is
such that there is one particle per cross sectional area $\sigma$.
This implies total particle number
\begin{equation}
\label{pn}
N ~=~ \pi\, r_{\perp}^2\,/\sigma ~,
\end{equation}
where $r_{\perp}$ must satisfy the inequality (\ref{causality}).

As an application of the above analysis, consider the simulation depicted
in Figure 2. We see well developed effects of the nonlinear interactions
by time $t = 600$.  This implies that a three dimensional wavepacket of
extent $r_{\perp} > 600$, which initially consists of $N = \pi\,
r_{\perp}^2\,/\sigma$ particles, will exhibit interesting nonlinear
interactions.

Of course, the extrapolation given above to three dimensions
is somewhat conservative, relying only on causality.
One might guess that for an initial wavepacket of the type depicted
in Figure 11, in which the transverse momenta of individual modes
is much smaller than their longitudinal momenta
$ k_{\perp} \ll k_{\,\parallel} $,
the nonlinearities can continue to develop even at times
after $t = r_{\perp}$. Individual modes will not disperse
appreciably in the transverse direction until a time
\begin{equation}
\label{td}
t_d ~\approx~ \frac{k_{\,\parallel}}{k_{\perp}}~ r_{\perp} ~.
\end{equation}
If this is the case, we should allow ourselves to follow
the one dimensional simulations  until time $t_d$, and conclude
that the corresponding three dimensional simulation will
behave similarly. Since $r_{\perp}$ can now be significantly
smaller than the run time of the simulation, we would predict
interesting nonlinear behavior for three dimensional wavepackets
consisting of fewer particles than estimated above.

\mysection{Conclusions}

In this paper, we have described the behavior of one dimensional
wavepacket collisions in pure Yang--Mills theory.  We have seen evidence
for the production of final states with dramatically different momentum
distributions, but for initial states with large average multiplicity.
When combined with the results of \cite{GHP}, these classical trajectories
imply unsuppressed quantum scattering amplitudes in Yang--Mills theory,
with initial wavepacket quantum states scattering to final states of very
different multiplicity.  It is worth recalling the situation which was
found in the study of multiparticle electroweak processes including those
with $(B+L)$ violation.  Certain processes with many particles (bosons)
both in the initial and in the final state ({\em many} $\rightarrow$
{\em many} scattering) can be shown to be unsuppressed within the
semiclassical approximation~\cite{KRT3}.  The methods of \cite{KRT3} are
Euclidean, and hence describe classically disallowed transitions, with the
suppression disappearing as the energy approaches the barrier where the
transitions are classically allowed.  Because our simulations are in
Minkowski space, they describe classically allowed processes.

Our simulations were performed with the coupling $g=1$.
By scaling the gauge potential, one can reach other values. For instance,
Figure 2 implies nonlinear behavior for $g=1$, $\sigma=1$, and initial state
with $N \approx 10^6$ particles (see (\ref{pn})).
Conversely a {\em two} particle collision, implying $\tilde{\sigma}\approx
10^6$ on a one dimensional lattice of the same size, would require a coupling
of $\tilde{g} = g \sqrt{\tilde{\sigma}}/\sqrt{\sigma}\approx 10^3$ to exhibit
similar nonlinear behavior (see (\ref{scaling})). For such a large value of the
coupling constant, the semiclassical approximation would not be expected to be
a good guide for the behavior of the quantum theory.
In the language of \cite{GHP},
there is no window in values of the coupling constant which yields
both interesting trajectories and controlled semiclassical corrections if
we apply our conservative criterion (\ref{causality}) for applicability to
three dimensions.

It is important to remember that the failure to find a trajectory
connecting a two particle initial state and a many particle
final state {\em does not} constitute
conclusive evidence that $2~\rightarrow$ {\em many}
amplitudes are suppressed.  Real Minkowski trajectories comprise only
a subset of the complex trajectories which may dominate multiparticle
scattering amplitudes at weak coupling~\cite{GHP}.

There are a few immediate directions for improving our results.

{\bf 1.}  We can  refine our study by including transverse sites on
the lattice.  An asymmetric lattice may allow this improvement without
drastically increasing the computation time.
As mentioned in Section 4, this would allow us to study
wavepackets with smaller transverse extent and fewer number of
particles.
We expect the one dimensional simulation to have determined an
upper bound on the critical number of particles in the three dimensional
world, so that we anticipate finding nonlinear effects for smaller numbers
of particles on the three dimensional lattice.

{\bf 2.} We can address the problem of multiparticle amplitudes in the
electroweak theory by including a Higgs scalar.
In this case, the explicit mass scale could act as a cut--off
on long wavelength excitations.
The additional possibilities of scalar--vector and scalar--scalar
scattering add to the complexity of the parameter space.

These issues will be investigated in future research.

\vspace{0.6in}

The authors would like to thank Krishna Rajagopal for useful comments
on the scaling behavior of our results.
This work was supported in part by the U.S. Department of Energy
(Grant No. DE-FG05-90ER40592) at Duke University and by a computing
grant from North Carolina Supercomputing Center.
S.G.M. would specially like to thank Professor Vartan Gregorian,
President of Brown University, and Office of International Programs of Duke
University for their support.
T.M.G. acknowledges the support of NSF grants \mbox{NSF-PHY-90-96198}
and \mbox{NSF-PHY-94-04057}.
S.D.H.H. acknowledges the support of DOE contract DE-AC02-ERU3075 at
Yale University.
E.R.P. is supported by a Robert R. McCormick fellowship and by
DOE contract DE-FGP2-90ER40560.

\newpage
\nc{\ib}[3]{        {\em ibid. }{\bf #1} (19#2) #3}
\nc{\np}[3]{        {\em Nucl.\ Phys. }{\bf #1} (19#2) #3}
\nc{\pl}[3]{        {\em Phys.\ Lett. }{\bf #1} (19#2) #3}
\nc{\pr}[3]{        {\em Phys.\ Rev.  }{\bf #1} (19#2) #3}
\nc{\prep}[3]{      {\em Phys.\ Rep.  }{\bf #1} (19#2) #3}
\nc{\prl}[3]{       {\em Phys.\ Rev.\ Lett. }{\bf #1} (19#2) #3}

\newpage

\centerline{\large\bf Figures}
\bigskip

Figure 1: Collision for parallel isospin polarization.
$\bar k = \pi/10$, $\Delta k = \pi/100$, and $\sigma = 1$. The wavepackets are
initially separated by $\Delta Z = 384$ on a lattice of length $L = 1024$ and
lattice spacing $a = 1$.

Figure 2: Same as Figure 1, but with orthogonal isospin polarization.

Figure 3: Time slices of energy spectrum (parallel isospin polarization).
$\bark = \pi/10$, $\Delta k = \pi/100$, and $\sigma = 1$. The units
for the momentum are $\pi/512$. The wavepackets are initially separated by
$\Delta Z = 384$ on a lattice of length $L = 1024$ and lattice spacing $a = 1$.

Figure 4: Same as Figure 3, but with orthogonal isospin polarization.

Figure 5: Amplitude dependence of instability shown in terms of different
$\sigma$ at time $t = 1200$ (orthogonal isospin polarization). Again we choose
$\bar k = \pi/10$ and $\Delta k = \pi/100$. This fixes the width parameter
$W=10$.
Different values for the amplitude parameter $R$ are obtained by varying
$\sigma$.
The units for the momentum are $\pi/1024$. The wavepackets are initially
separated by $\Delta Z = 1024$ on a lattice of length $L = 2048$ and lattice
spacing $a = 1$.

Figure 6: Fixed time slices of energy spectrum which show
dependence of the instability on  the dimensionless
width of the wavepacket $W$ with dimensionless amplitude $R$ fixed.
All time slices are at $t = 800$ with $\bark=\pi/10, a=1, L=2048$ and
$\Delta Z=1024$. The units for the momentum are $\pi/1024$.

Figure 7: Same as Figure 6, but with $t = 1200$.

Figure 8: Spectra at $t=600$ for different isospin polarization $\theta_c$.
$\bar k = \pi/10$, $\Delta k = \pi/100$, and $\sigma = 1$. The units for
the momentum are $\pi/512$. The wavepackets are initially separated by $\Delta
Z = 384$ on a lattice of length $L = 1024$ and lattice spacing $a = 1$.

Figure 9: Energy contained in the isospin $c=1$ component of the gauge field,
shown as a function of time in terms of different $\theta_c$. $\bar k =
\pi/10$, $\Delta k = \pi/100$, and $\sigma=1$. The wavepackets are initially
separated by $\Delta Z = 384$ on a lattice of length $L = 1024$ and lattice
spacing $a = 1$.

Figure 10: The edge of the wavepacket at $r_\perp$ is out of causal contact
with the interior region for sufficiently short times $t$.

Figure 11: Collision of two wavepackets in coordinate space.

\newpage
\ifx\pansw\pictures
\def\epsfsize#1#2{0.95#1}
\centerline{\epsfbox{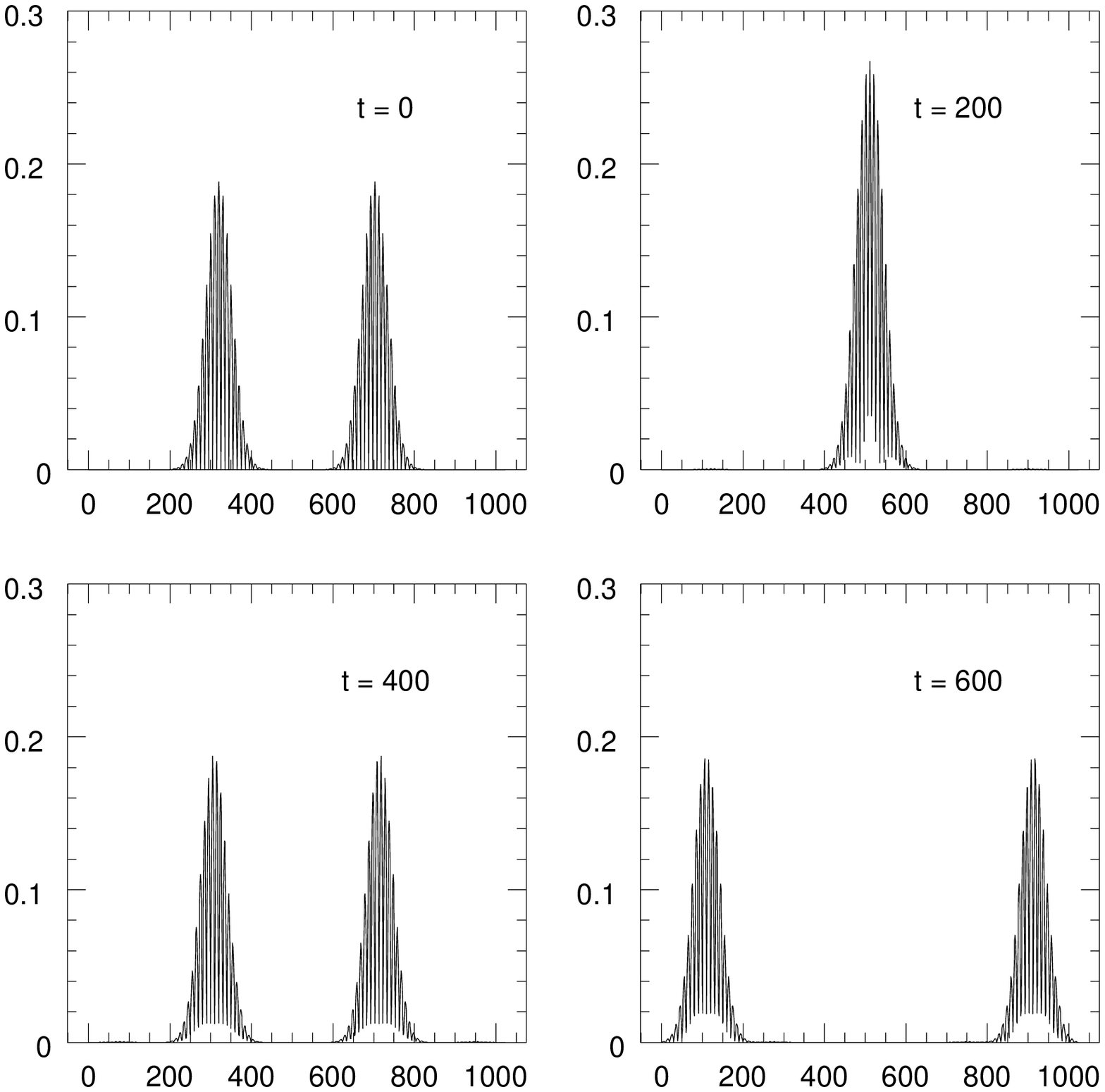}}
\centerline{Fig.1}
\else\fi

\ifx\pansw\pictures
\def\epsfsize#1#2{0.95#1}
\centerline{\epsfbox{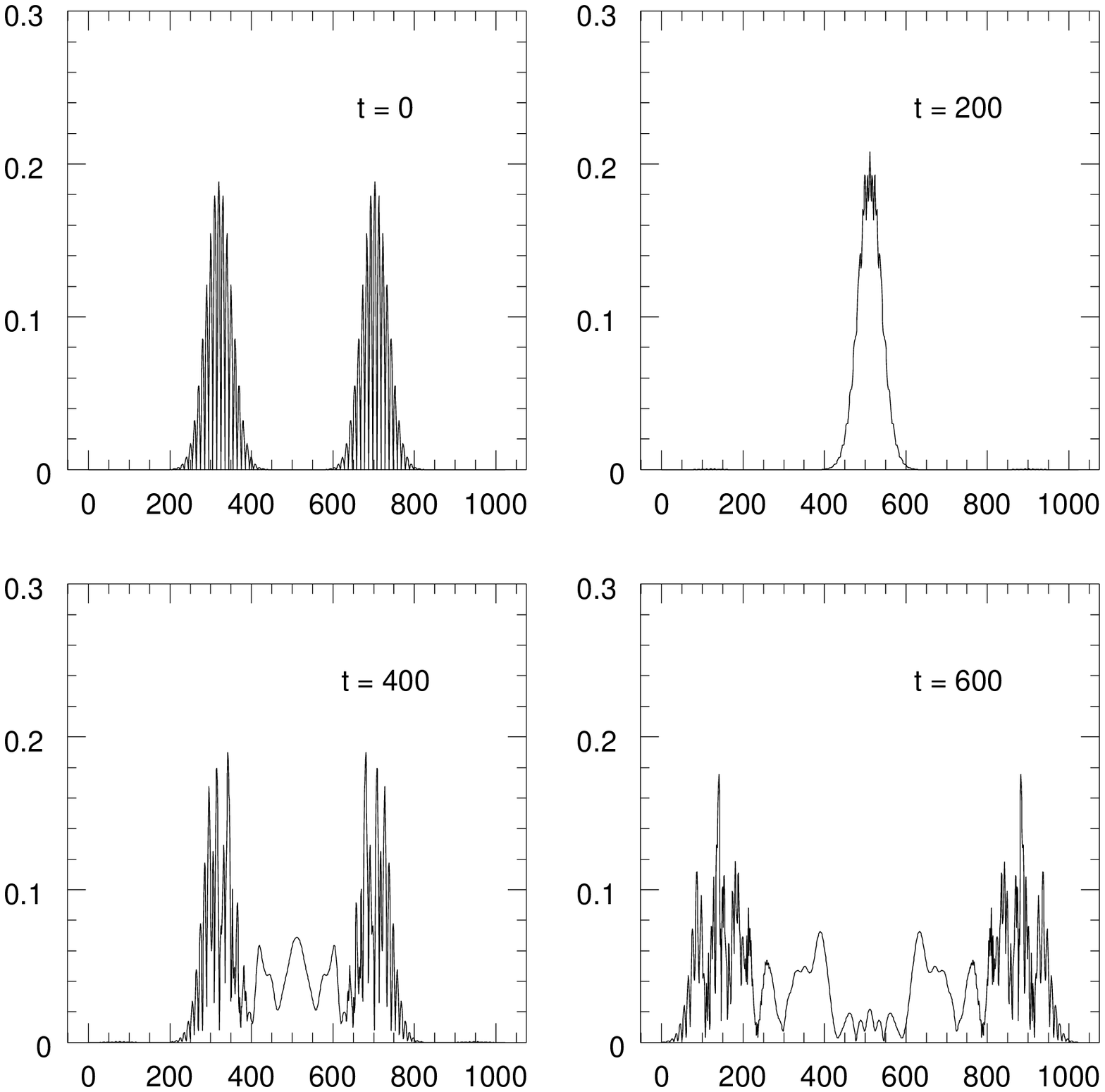}}
\centerline{Fig.2}
\else\fi

\ifx\pansw\pictures
\def\epsfsize#1#2{0.95#1}
\centerline{\epsfbox{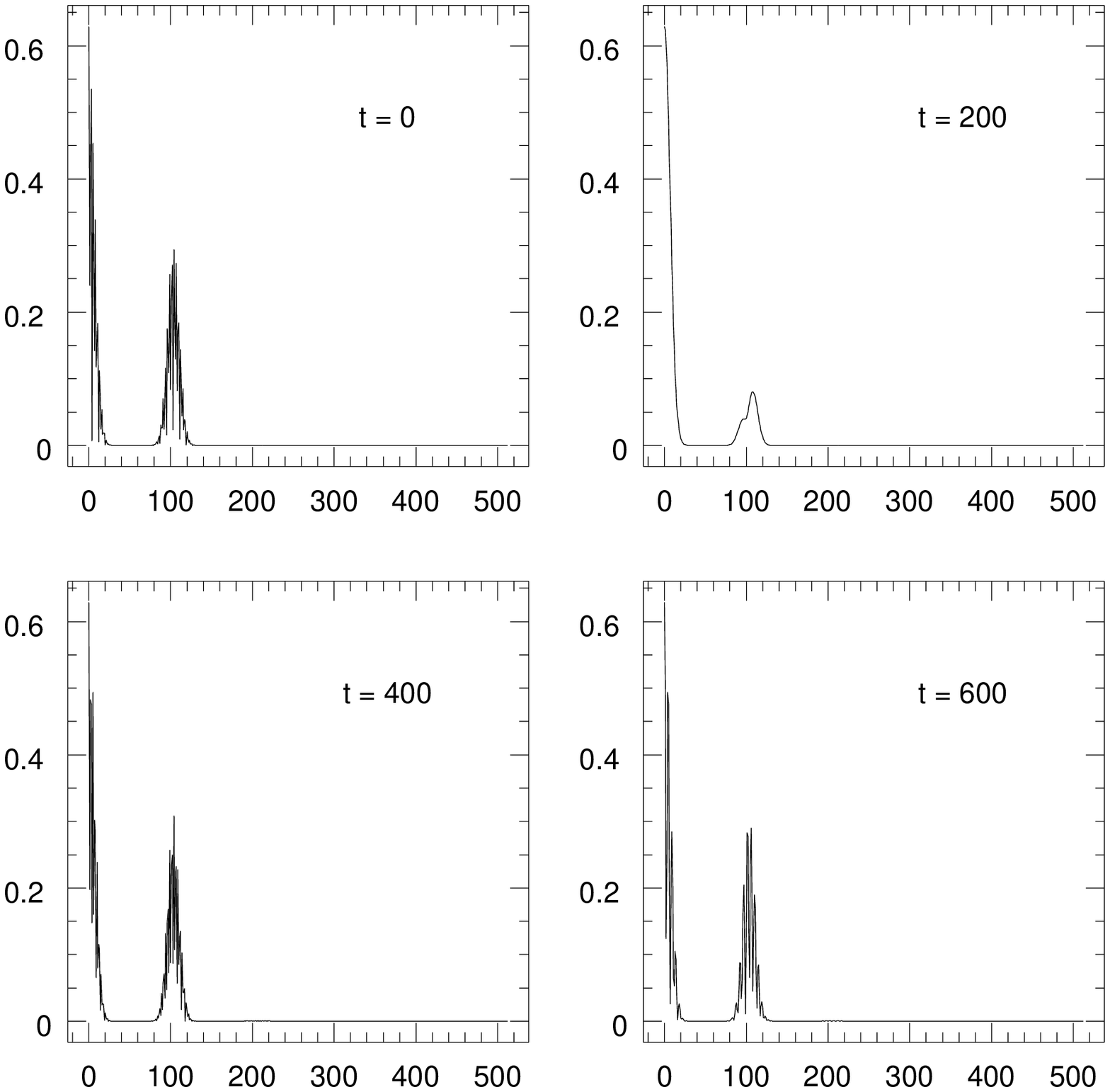}}
\centerline{Fig.3}
\else\fi

\ifx\pansw\pictures
\def\epsfsize#1#2{0.95#1}
\centerline{\epsfbox{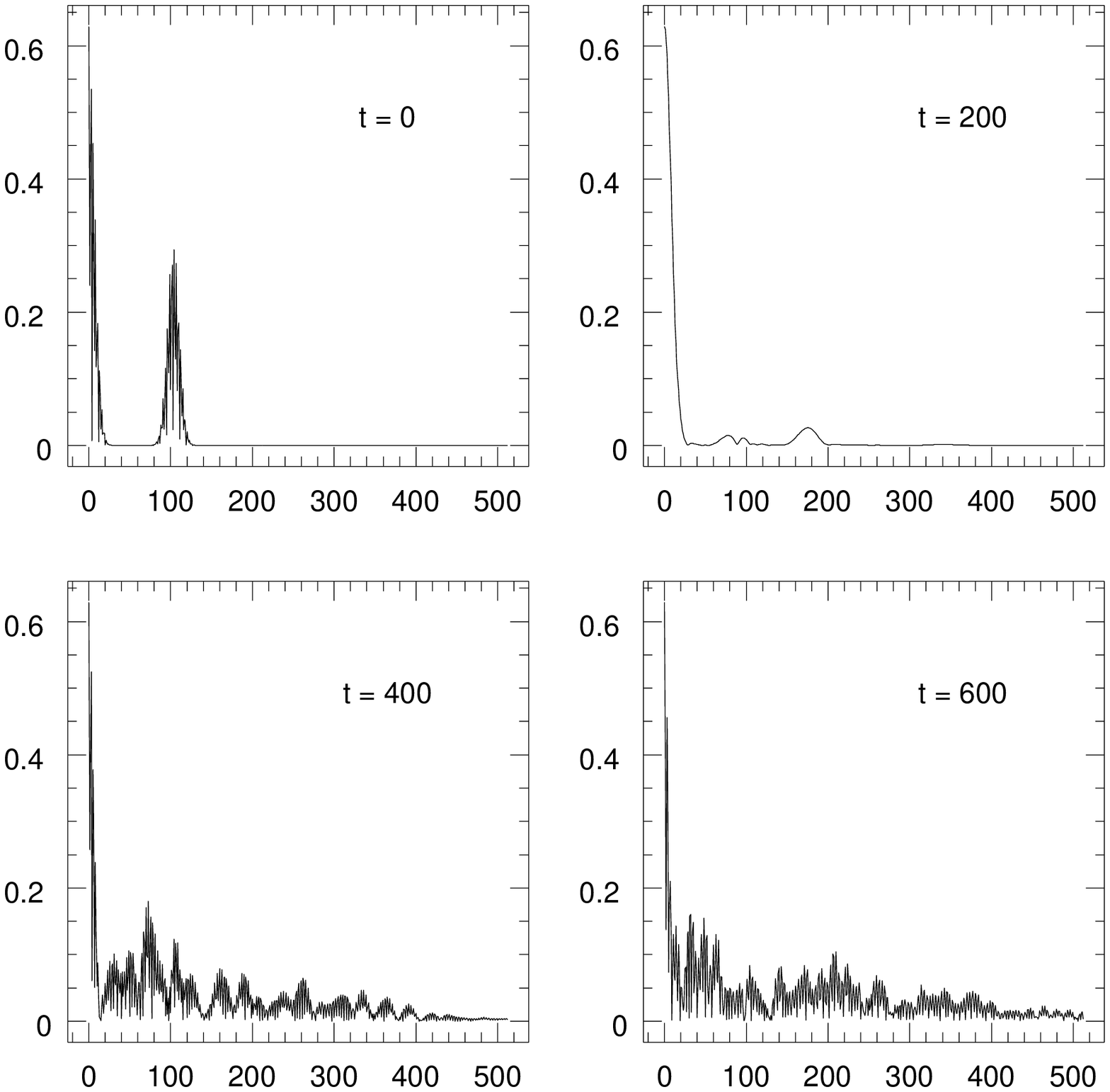}}
\centerline{Fig.4}
\else\fi

\ifx\pansw\pictures
\def\epsfsize#1#2{0.95#1}
\centerline{\epsfbox{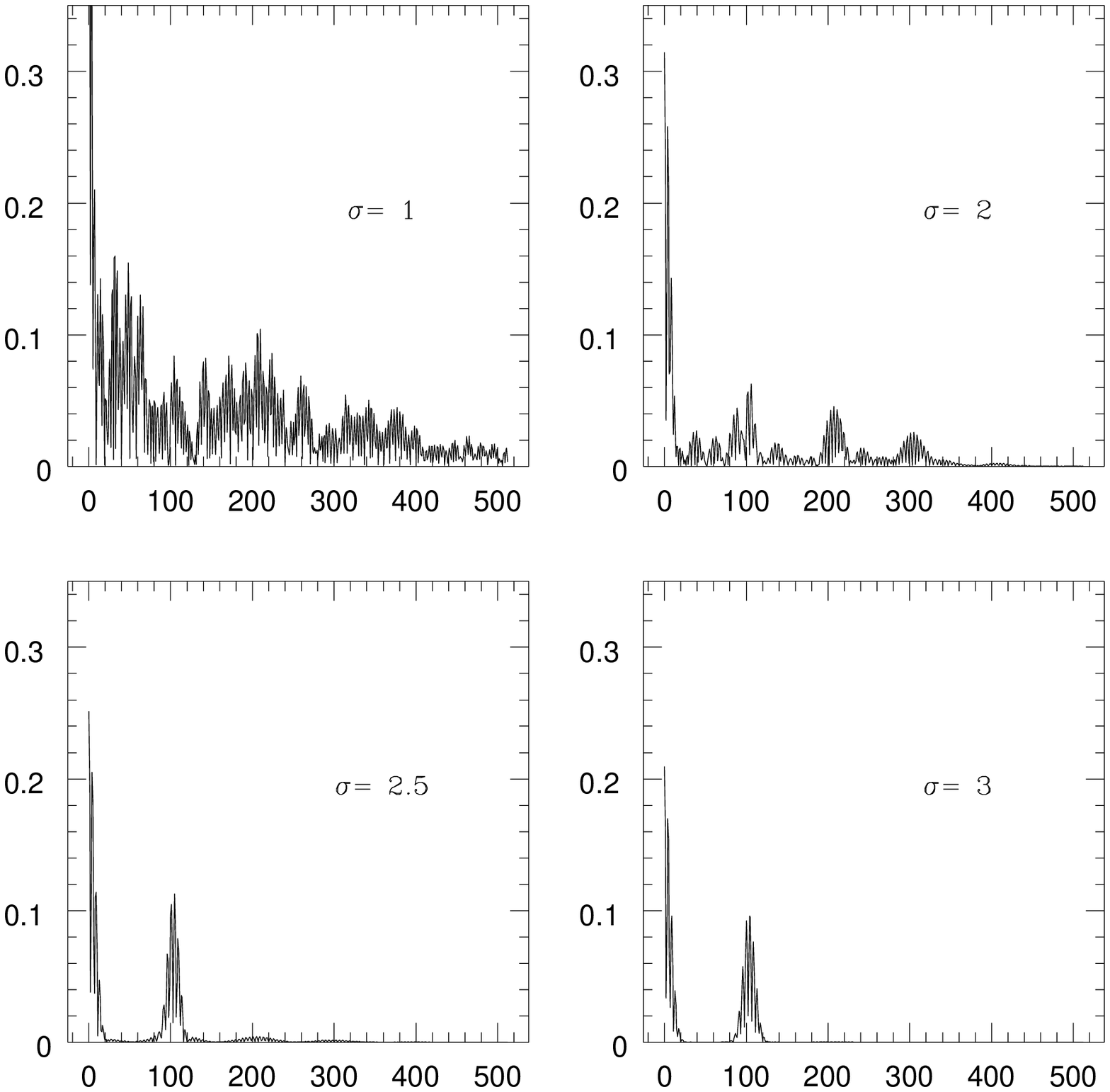}}
\centerline{Fig.5}
\else\fi

\ifx\pansw\pictures
\def\epsfsize#1#2{0.95#1}
\centerline{\epsfbox{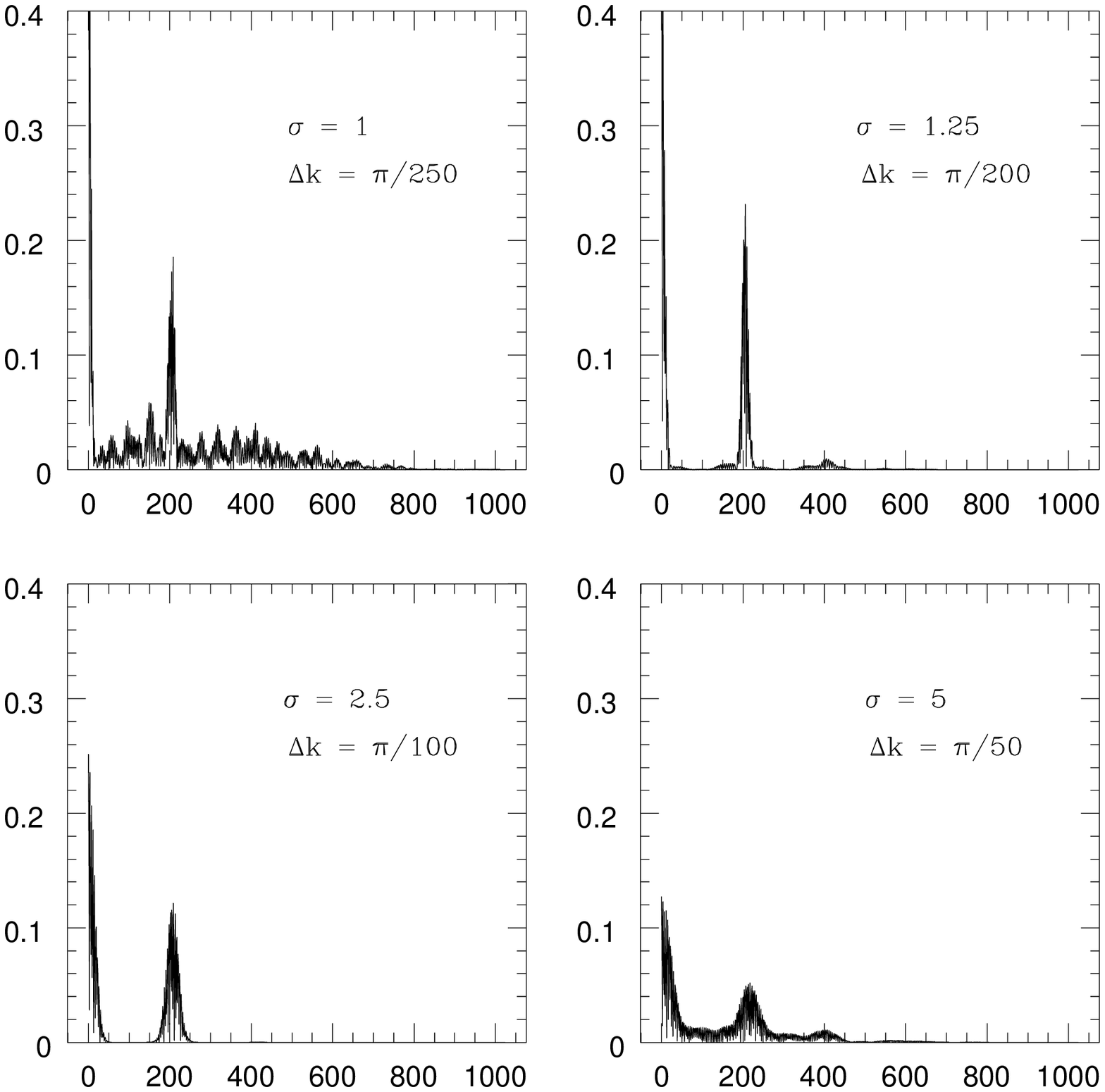}}
\centerline{Fig.6}
\else\fi

\ifx\pansw\pictures
\def\epsfsize#1#2{0.95#1}
\centerline{\epsfbox{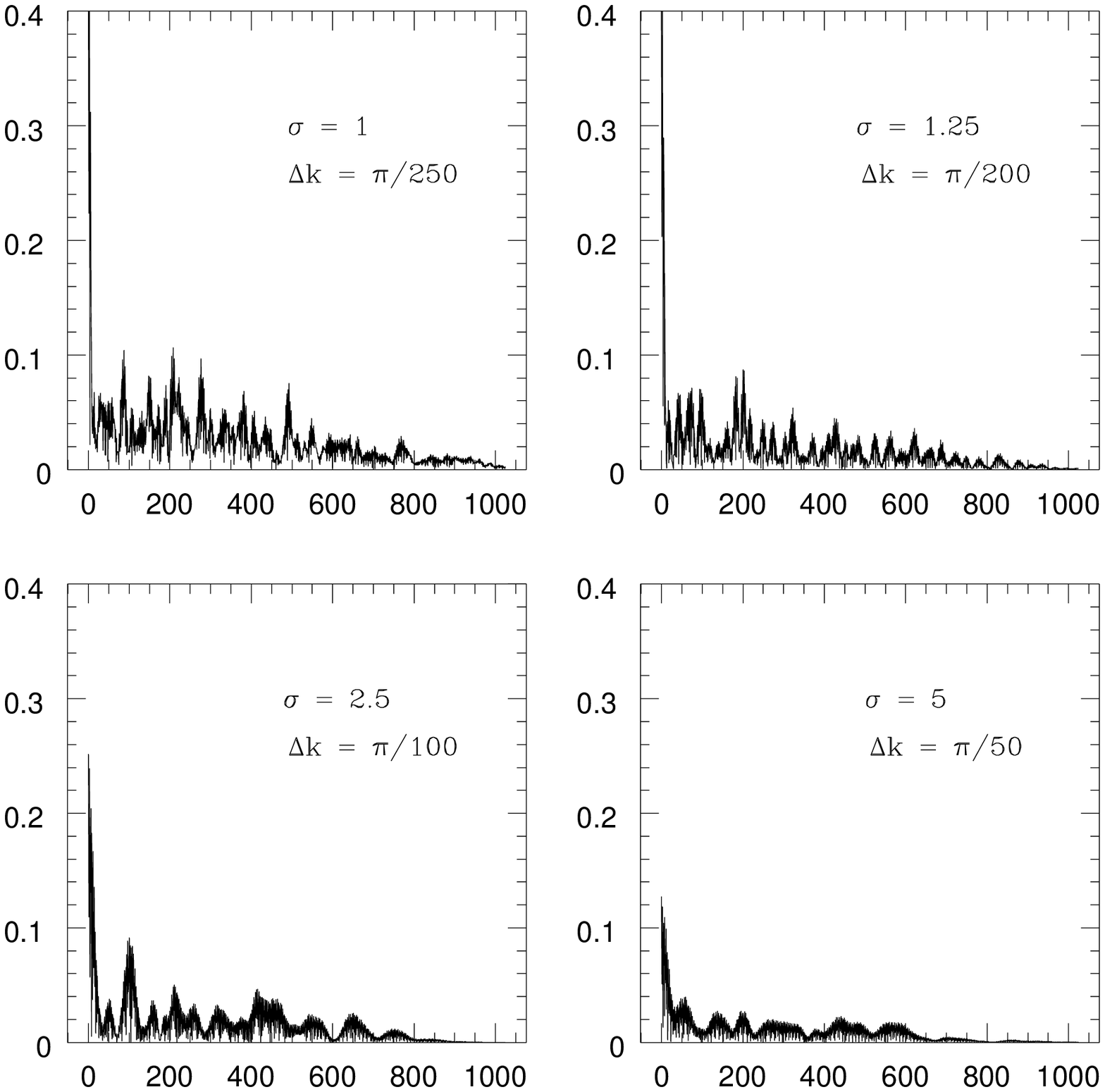}}
\centerline{Fig.7}
\else\fi

\ifx\pansw\pictures
\def\epsfsize#1#2{0.95#1}
\centerline{\epsfbox{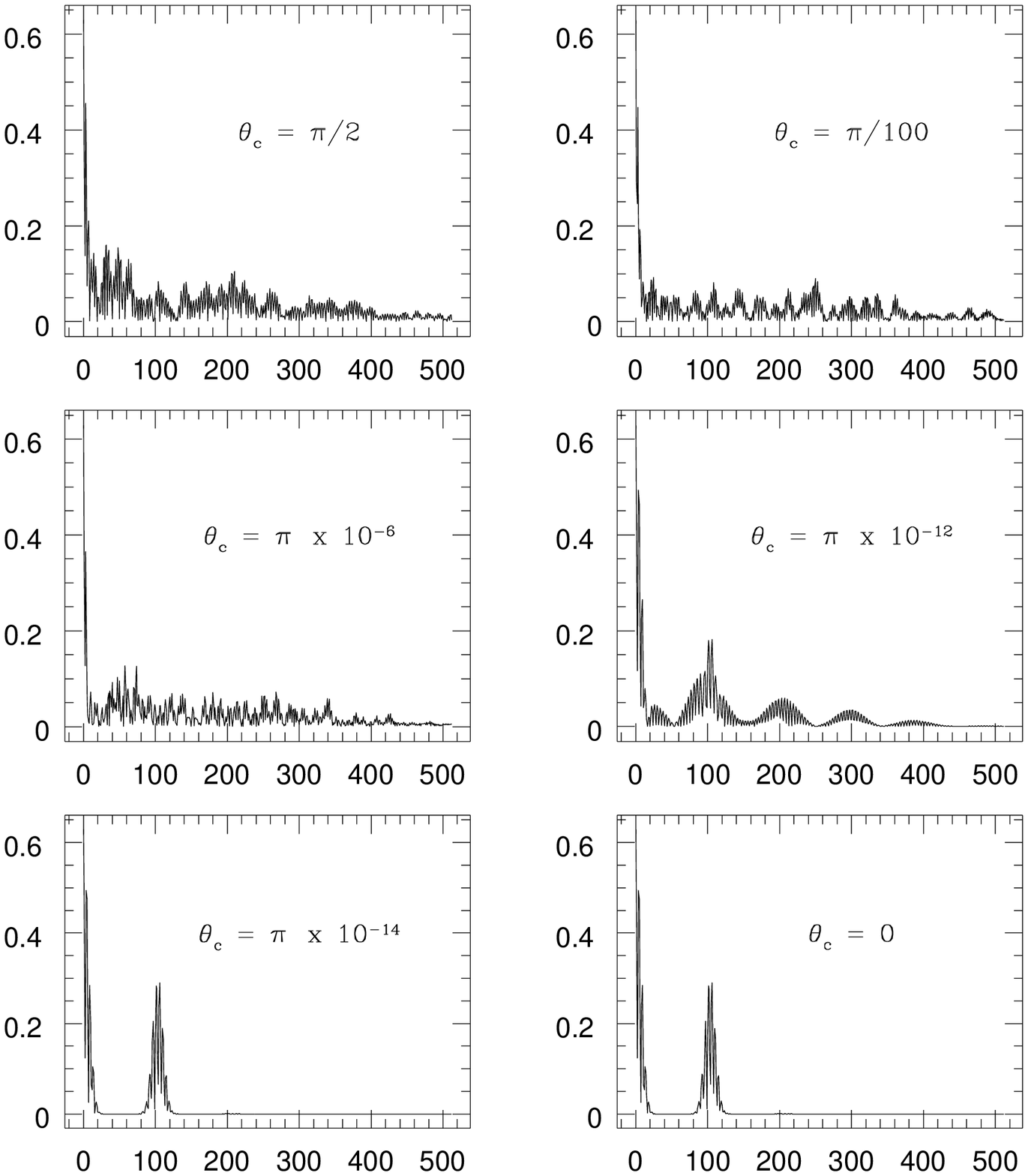}}
\centerline{Fig.8}
\else\fi

\ifx\pansw\pictures
\def\epsfsize#1#2{0.95#1}
\centerline{\epsfbox{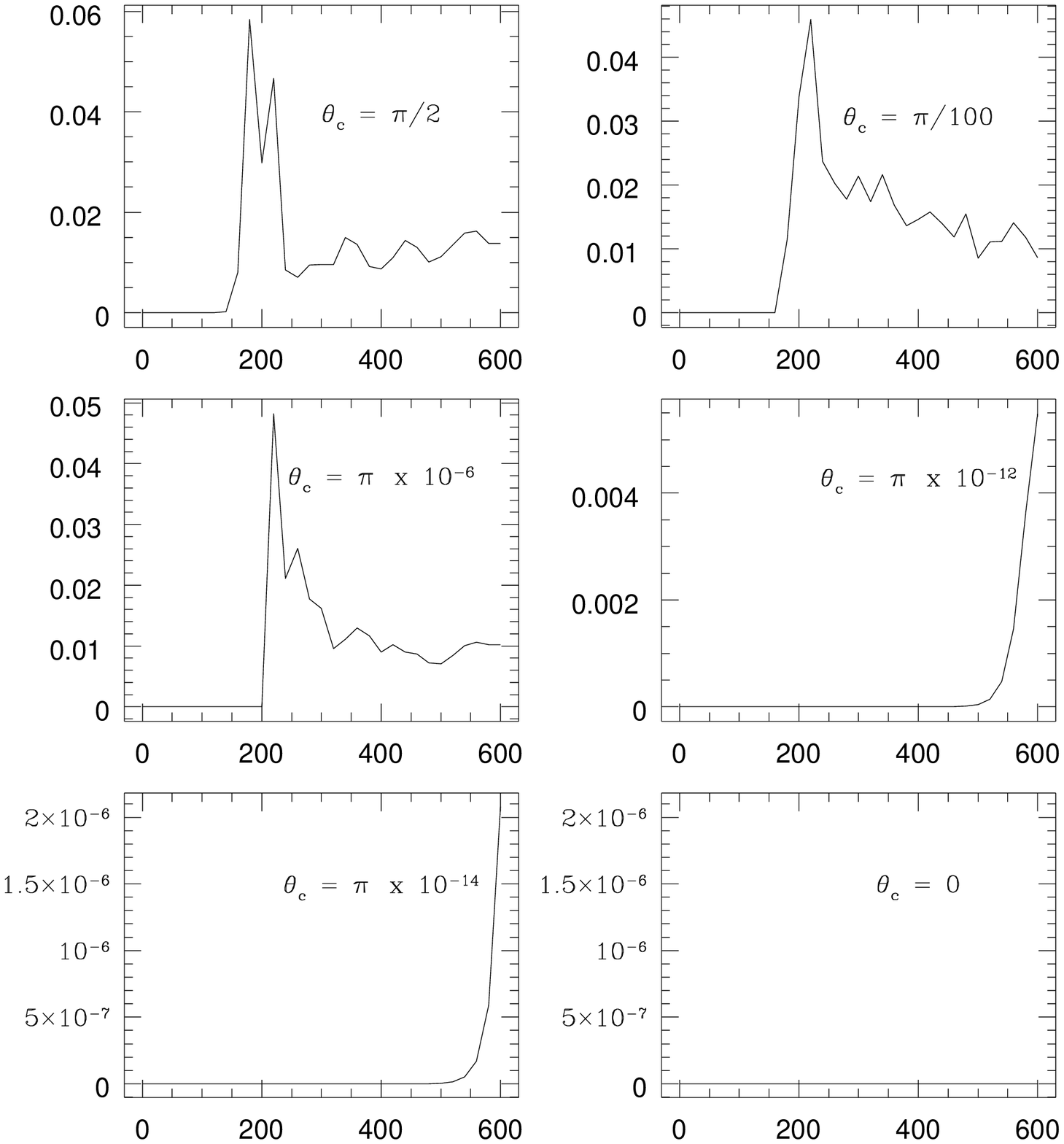}}
\centerline{Fig.9}
\else\fi

\newpage
\ifx\pansw\pictures
$$
\beginpicture
\setplotarea x from -65 to 65 , y from -40 to 55
\arrow < 4 pt> [0.4,1.5] from 0 0 to 0 140  \put{$t$} at 0 150
\arrow < 4 pt> [0.4,1.5] from -100 40 to 140 40  \put{$~r$} at 150 40
\arrow < 4 pt> [0,0] from 100 40 to 160 100
\arrow < 4 pt> [0,0] from 100 40 to  30 110
\putrule from 100 38 to 100 42 \put{$r_\perp$} at 100 30
\setdots
\putrule from -100 100 to 100 100  \put{$t<r_\perp$} at -120 100
\setshadegrid span <2pt> \shaderectangleson
\putrectangle corners at -40 40 and 40 100 \setsolid
\put{ Fig.10}
at 0 -50
\endpicture
$$
\bigskip
\else\fi

\newpage
\ifx\pansw\pictures
$$
\beginpicture
\setplotarea x from -65 to 65 , y from -40 to 55
\ellipticalarc axes ratio 1:3 360 degrees from -60 50 center at -70 50
\setshadegrid span <\tdim>
\vshade -80 40 60 <,z,,> -78 35 65  <,z,,>
-76 28 72 <,z,,> -74 22 78 <,z,,>  -72 20 80 <,z,,>
-70 18 82 <,z,,> -68 20 80 <,z,,> -66 22 78 <,z,,>
-64 28 72 <,z,,> -62 35 65 <,z,,> -60 40 60 /
\ellipticalarc axes ratio 1:3 360 degrees from  60 50 center at  70 50
\setshadegrid span <\tdim>
\vshade 60 40 60 <,z,,> 62 35 65  <,z,,>
64 28 72 <,z,,> 66 22 78 <,z,,>   68 20 80 <,z,,>
70 18 82 <,z,,> 72 20 80 <,z,,> 74 22 78 <,z,,>
76 28 72 <,z,,> 78 35 65 <,z,,> 80 40 60 /
\arrow < 4 pt> [0.4,1.5] from -50 40 to -30 40
\arrow < 4 pt> [0.4,1.5] from -50 60 to -30 60
\arrow < 4 pt> [0.4,1.5] from  50 40 to  30 40
\arrow < 4 pt> [0.4,1.5] from  50 60 to  30 60
\arrow < 4 pt> [0.4,1.5] from 0 100 to 0 120  \put{$r$} at 0 130
\arrow < 4 pt> [0.4,1.5] from 100 50 to 120 50 \put{$z$} at 130 50
\put{ Fig.11}
at 0 -30
\endpicture
$$
\bigskip
\else\fi

\end{document}